\documentclass[a4paper,fleqn,usenatbib]{mnras}

\usepackage{newtxtext,newtxmath}

\usepackage[T1]{fontenc}
\usepackage{ae,aecompl}

\usepackage{graphicx}	
\usepackage{amsmath}	
\usepackage{pdflscape}
\usepackage{soul}
\usepackage{xcolor,colortbl}
\usepackage{float}

\title[Modeling MHD Equilibrium in Magnetars]{Modeling Magnetohydrodynamic Equilibrium in Magnetars with Applications to Continuous Gravitational Wave Production}

\author[S.G Frederick et al.]{
S. G. Frederick$^{1}$,\thanks{E-mail: safrederick@davidson.edu}
K. L. Thompson$^{1}$,
M. P. Kuchera$^{1}$
\\
$^{1}$Department of Physics, Davidson College, 405 N Main St, Davidson, NC 28035, USA
}

\begin{document}
\label{firstpage}
\pagerange{\pageref{firstpage}--\pageref{lastpage}}
\maketitle

\begin{abstract}
Possessing the strongest magnetic fields in the Universe, magnetars mark an extremum of physical phenomena. The strength of their magnetic fields is sufficient to deform the shape of the stellar body, and when the rotational and magnetic axes are not aligned, these deformations lead to the production of gravitational waves (GWs) via a time-varying quadrupole moment. Such gravitational radiation differs from signals presently detectable by the Laser Interferometer Gravitational-Wave Observatory. These signals are continuous rather than the momentary `chirp' waveforms produced by binary systems during the phases of inspiral, merger, and ringdown. 
Here, we construct a computational model for magnetar stellar structure with strong internal magnetic fields. We implement an $n = 1$ polytropic equation of state (EOS) and adopt a mixed poloidal and toroidal magnetic field model constrained by the choice of EOS. We utilize fiducial values for magnetar magnetic field strength and various stellar physical attributes. Via computational simulation, we measure the deformation of magnetar stellar structure to determine upper bounds on the strength of continuous GWs formed as a result of these deformations inducing non-axisymmetric rotation. We compute predictions of upper limit GW strain values for sources in the McGill Magnetar Catalog, an index of all detected magnetars. 
\end{abstract}




\section{Introduction}
Magnetars are an exceptional classification of pulsars, characterized by surface magnetic field strengths in excess of $10^{14}$ G and dipolar magnetic energies exceeding the star's rotational energy \citep{Thompson_Duncan_1995}. \citet{mcgill} provide a catalog of 23 confirmed and 6 candidate sources, and document considerable progress in magnetar detection via $\gamma$-ray burst events in recent years following the launch of the \textit{Swift} and \textit{Fermi} space telescopes. Given the rapid growth in confirmed magnetar sources, these stars present a wealth of opportunity for improving current understanding regarding the influence of strong magnetic fields in extreme stellar environments.   


 \citet{chandrasekhar_fermi_1953} first showed for an incompressible stellar model that a strong internal magnetic field will deform a star away from spherical symmetry. For deformations induced along a magnetic field axis which is misaligned with the stellar rotational axis, a time-varying gravitational quadrupole will result in the production of gravitational waves (GWs). Thus, magnetars are compelling candidates for the detection of GWs from deformed stellar sources. 

Such GWs differ from former event detections, as unlike the `chirp' waveform of binary inspiral mergers, GWs produced by a rapidly rotating stellar source are nearly constant-frequency, sinusoidal signals due to the source returning to the same spatial configuration in the span of a complete revolution about its rotational axis. Due to the consistent periodicity of these GW signals, they are referred to as `continuous' GWs. Under extended survey, stellar spin down due to loss in rotational kinetic energy through magnetic braking or energy loss in the form of gravitational radiation will increase the rotational period and GWs emitted will drift to lower frequencies \citep{creighton_anderson_2011}. However, under shorter observation, continuous GWs appear as constant frequency sinusoidal waveforms. 

 Continuous GWs are expected to be detected following improvements in GW detector sensitivity, as their signals are often far fainter than GWs produced by binary inspiral events. Evaluation of their signal strength, or wave strain, can be made by estimating the magnitude of stellar deformations responsible for producing such signals \citep{zimmermann_szedenits_1979}. 

Recent work places upper limits on the GW strain of pulsar sources capable of producing GWs within the operating range of the Laser Interferometer Gravitational-Wave Observatory (LIGO) \citep{Abbott_etal_2017}. The authors compute the spin-down limit; the GW strain sensitivity produced by attributing the loss in rotational kinetic energy completely to gravitational radiation. For a rigidly rotating triaxial star, the frequency of gravitational waves produced by the source will be twice the rotational frequency. As magnetars are slowly rotating stars (with rotational periods $\sim 5$--$8$ s) \citep{mcgill}, GWs produced by these sources fall outside the sensitivity range of LIGO and corresponding wave strain estimates were not addressed by \citet{Abbott_etal_2017}. 

The principal goal of this paper is to provide estimates for upper-limit calculations of the GW strain for all confirmed magnetar sources in the McGill Magnetar Catalog \citep{mcgill} by constructing a computational model for magnetar stellar structure and magnetic field configuration. We determine the degree of structural deformation introduced by a strong internal magnetic field as the stellar structure reaches magnetohydrodynamic (MHD) equilibrium. These results for stellar deformation subsequently inform wave strain estimates.


To compute upper-limit estimates of the GW strain for magnetars, we adopt a simple barotropic EOS for an n=1 polytrope and seek dipolar solutions to the mixed poloidal-toroidal magnetic field model derived by \citet{haskell_2008}. We then compute the ellipticity that arises from this solution and calculate the GW strain for the magnetars in the McGill catalog.

For numerically stable computation, we use the fiducial values for stellar attributes: stellar mass, $M_{\star} = 1.4$ M$_{\odot}$; stellar radius, $R_{\star} = 10$ km; and central density, $\rho_c = 2.2\times10^{15}$ g$\cdot$cm$^{-3}$. 

\section{Structural Model}
Prior authors \citep[][and references therein]{Owen} note that cumulative errors introduced by excluding relativistic gravity and rotational effects largely cancel; while relativistic gravity results in a more compact model of stellar structure than the Newtonian framework, stellar rotation has an opposing effect. Thus, in constructing a stellar model, we adopt the Newtonian gravitational theory and neglect rotational effects.  

This gives us the following set of non-relativistic MHD equations to describe the time evolution of the system.
\begin{subequations}
\begin{align}
    \frac{\partial \rho}{\partial t} + {\bf v} \cdot \nabla \rho + \rho \nabla \cdot {\bf v} & = 0\\
    \frac{\partial {\bf v}}{\partial t} + {\bf v} \cdot \nabla {\bf v} + \frac{1}{\rho}{\bf B} \times (\nabla \times {\bf B}) + \frac{1}{\rho}\nabla \rho & = -\nabla \Phi + {\bf g} \\
    \frac{\partial {\bf B}}{\partial t} + {\bf B}(\nabla \cdot {\bf v})  - ({\bf B \cdot \nabla}){\bf v} + ({\bf v} \cdot \nabla){\bf B } &= {\bf v(\nabla \cdot B)} \\
    \frac{\partial p}{\partial t} + {\bf v} \cdot \nabla p + \rho c_s^2 \nabla \cdot {\bf v} & = 0
\end{align}
   \label{eq:newt}
\end{subequations}
where {\bf v} is the  velocity vector, {\bf B} is the magnetic field, $\rho$ is density, $p$ is momentum, {\bf g} is the gravitational acceleration vector, $\Phi$ is the time-independent gravitational potential,  and $c_s$ is the adiabatic speed of sound  as described in  \cite{PLUTO_userguide}. All of the computations in this paper evolve these sets of equations numerically using the third-order Runge Kutta algorithm for time evolution.

\subsection{Hydrostatic Equilibrium Conditions}
Our choice of stellar model is constrained to configurations which are in equilibrium. Thus, the construction of this stellar model requires a crucial balance between the force of gravity and stellar structure. The equilibrium condition
\begin{equation}
    \frac{dP}{dr} = -G\frac{M_r\rho}{r^2},
\end{equation}
where $P$ and $\rho$ are the stellar pressure and density, respectively, $G$ is the gravitational constant, and $M_{r}$ is the mass interior to the radius for $r< R_{\star}$,  provides the basis for balancing the gravitational force with structural variation throughout the stellar interior.

The interior mass varies with radius, and thus introduces the following equation for mass conservation within the stellar medium:
\begin{equation}
    \frac{dM_r}{dr} = 4\pi r^2 \rho. 
\end{equation}

\subsection{Polytropic Equations of State}
The time-independent equations of hydrostatic equilibrium and mass conservation provide an initial description of stable Newtonian stars. In order to fully specify stellar structure, an equation of state (EOS) is required to relate pressure to a number of state variables describing stellar structure. We adopt a barotropic EOS, which defines the relationship between pressure and density as $P(\rho)$. While an EOS parameterized by numerous state variables such as $P(\rho,x_p,T,...)$ is more physically representative of the interior of a neutron star, subsequent discussion will show that our particular choice of barotropic EOS allows analytic equations for stellar structure and magnetic field. We use a polytropic EOS of the form
\begin{equation}
    P(\rho) = K\rho^{\gamma},
    \label{eq:polytrope}
\end{equation}
where $K$ is the polytropic constant and the real, positive constant $\gamma$ is defined via the polytropic index $n$ as
\begin{equation}
    \gamma = \frac{n+1}{n}.
    \label{eq:gamma}
\end{equation}


Polytropic equations of state are often categorized by the compressibility of stellar matter, whereby a lowering of the constant $\gamma$ corresponds to lower compression \citep{Haensel}. Thus, the structural composition of the stellar interior sets a constraint on representative equations of state. 

Prior work has established neutron star structure as well approximated by the choice of polytropic EOS corresponding to  $0<n\lesssim1$ (\cite{cho_lee_2010}, \cite{woosley_2014}).  In addition, an $n=1$ polytrope has the property that the stellar radius is unaffected by mass nor central density, which reflects the insensitivity of radius to mass within normal neutron stars. Under these considerations, we implement an $n=1$ polytropic EOS in modeling neutron star structure. An equation for density as a function of radius can be determined via solutions to the Lane-Emden equation for a specified polytropic index, $n$. The $n=1$ polytrope possesses the following analytic solutions for $\rho(r)$ and $P(r)$ from which stellar structure can be fully determined:
\begin{equation}
    \rho(r) = \rho_c \frac{\sin{(\pi r/R_{\star})R_{\star}}}{r\pi}\text{ \hspace{0.5cm} for $r<R_{\star}$}
    \label{eq:rho_eqn}
\end{equation}
and 
\begin{equation}
    P(r) = K\rho(r)^2\text{ \hspace{0.5cm} for $r<R_{\star}$},
    \label{eq:prs_eqn}
\end{equation}
where the polytropic constant $K=4.25 \times 10^4 \text{ cm}^5\cdot\hspace{-.1cm}\text{ g}^{-1}\cdot\hspace{-.1cm}\text{ s}^{-2}$ for a neutron star with radius $R_{\star}=10$ km, mass $M_{\star} = 1.4$ $M_{\odot}$, and central density $\rho_c = 2.2\times10^{15}$ g$\cdot$cm$^{-3}$. 

\subsection{Gravitational Potential Model}
Hydrostatic equilibrium requires the balance of an inward gravitational force with the radial change in pressure. We determine solutions to the spherically symmetric form of Poisson's equation for gravitational potential per unit mass,  
\begin{equation}
    \frac{1}{r^2}\frac{d}{dr}\left(r^2\frac{d\Phi_g}{dr}\right) = 4\pi G\rho.
    \label{eq:poisson}
\end{equation}
  Solutions to $\Phi_g$ interior and exterior to the stellar surface are constrained by density as given by Equation \ref{eq:rho_eqn}. Additionally, these solutions must satisfy the following boundary conditions: 
  \begin{subequations}
  \begin{align}
    \frac{d\Phi_g}{dr} &= 0\hspace{.1cm} \bigg\rvert_{r = 0},\\
    \Phi_g\hspace{.1cm}\bigg\rvert_{r = R_{\star}^{\text{ inside}}} &= \Phi_g\hspace{.1cm}\bigg\rvert_{r = R_{\star}^{\text{ outside}}},\\
    \frac{d\Phi_g}{dr}\hspace{.1cm}\bigg\rvert_{r = R_{\star}^{\text{ inside}}} &= \frac{d\Phi_g}{dr}\hspace{.1cm}\bigg\rvert_{r = R_{\star}^{\text{ outside}}}.
  \end{align}
  \end{subequations}

Using separation of variables and substitution to solve Equation \ref{eq:poisson} for $\Phi_g(r)$, we determine the following equations for the gravitational potential:
  \begin{subequations}
  \begin{align}
    \Phi_g^{\text{core}} &= 4G\rho_c\left(-\frac{R_{\star}^2}{\pi} - \frac{M}{4R_{\star}\rho_c}\right)\text{, \hspace{.03cm}for } r = 0\\
    \Phi_g^{\text{inside}}(r) &= 4G\rho_c\left(\frac{-R_{\star}^3\sin{(\pi r/R_{\star})}}{\pi^2 r} - \frac{M}{4R_{\star}\rho_c}\right)\text{, \hspace{.03cm}for } 0 < r < R_{\star}\\
    \Phi_g^{\text{outside}}(r) &= -\frac{GM}{r}\text{, \hspace{.03cm}for } r > R_{\star}.
  \end{align}
  \label{eq:gpot}
  \end{subequations}

 As the stellar model evolves and the magnetic field induces morphological changes in the stellar structure, the model's gravitational potential remains static; the potential adheres to the form determined here as assigned via initial conditions of the simulation. This approximation is referred to as the Cowling approximation, and provides considerable accuracy under direct comparison tests between static and dynamic potentials for modeling stellar structure which, under evolution, become slightly perturbed from initial conditions (\cite{yoshida_kojima_1997, yoshida_2013}).

\subsection{Computational Modeling}
We use the astrophysical fluid dynamics code PLUTO of \cite{PLUTO} to specify
and simulate computational stellar models. PLUTO allows users to specify a
computational domain for modelling fluid dynamics simulations under a variety of
physical scenarios including hydrodynamic (HD) and magnetohydrodynamic (MHD) flow.
Additionally, special relativistic effects can be modeled for both HD and MHD scenarios.
Simulated physical scenarios within the PLUTO code correspond to physics modules
available to the user. We use non-relativistic state equations due to our choice of Newtonian structural equations and gravitational potential to model stellar structure. PLUTO then evolves the equations in \ref{eq:newt} for simulating stellar structure with
and without a magnetic field present, respectively.
The physical module and other components of the simulation including parameters for
the geometry and dimensions of computational domain, forces within the simulation, and
other optionally simulated physical phenomena are specified in the source code of this project, which is available via \cite{frederick_github}. 
The configuration details of our simulation are presented in Table~\ref{tab:config}.

The computational domain is a three-dimensional, spherical region and the stellar model is centered
within the domain, extending radially to 1.1 times unit radius, where \texttt{R\_star = 1.0}. The
domain is radially discretized into 100 grid cells, and the azimuthal and polar axes are
modeled from (0,2$\pi$) and (0, $\pi$), respectively, and an angular resolution of 30 grid cells
is selected for both of these axes.
A third-order total variation diminishing Runge-Kutta scheme (\texttt{TIME\_STEPPING: RK3}) is
used for time stepping, and reconstruction of the simulation is achieved via a third-order
weighted essentially non-oscillatory scheme (\texttt{RECONSTRUCTION: WENO3}).

To ensure that our magnetic field model adheres to Maxwell's equations through the time evolution of Equations \ref{eq:newt}, particular significance is given to preserving the divergence condition, $\nabla\cdot\boldsymbol{B} = 0$,
for the purposes of accurate magnetic field evolution using hyperbolic divergence cleaning  \citep{2002JCoPh.175..645D}.
We choose the Hartman, Lax, Van Leer contact method (hllc) as our Riemann solver for integration across our computational grid, as it is shown in \cite{hllc} to be suitable in MHD calculations such as the ones used in this paper.

The boundary conditions for our computational domain are outlined in Table~\ref{tab:BC}.

The outflow condition sets zero gradient across the interior radial boundary.  Axisymmetric boundary conditions for the
polar boundaries reflect vector valued-variables across the boundary. Periodic boundary
conditions repeat vector-valued variables across both azimuthal boundaries.


\subsection{Structural Validation}
We test the equilibrium condition of 
equilibrium 
for our stellar model by generating a 3-D spherical computational simulation of an $n=1$ polytrope with unit radius $R_{\star}=1$ in dimensionless code units and accompanying gravitational potential in the form of Equation \ref{eq:gpot}. The computational simulation is composed of finite-volume voxels, each assigned structural parameter values interpolated via the analytic form of Equations \ref{eq:rho_eqn}, \ref{eq:prs_eqn}, and \ref{eq:gpot}.

\begin{figure}
    \includegraphics[width=\columnwidth]{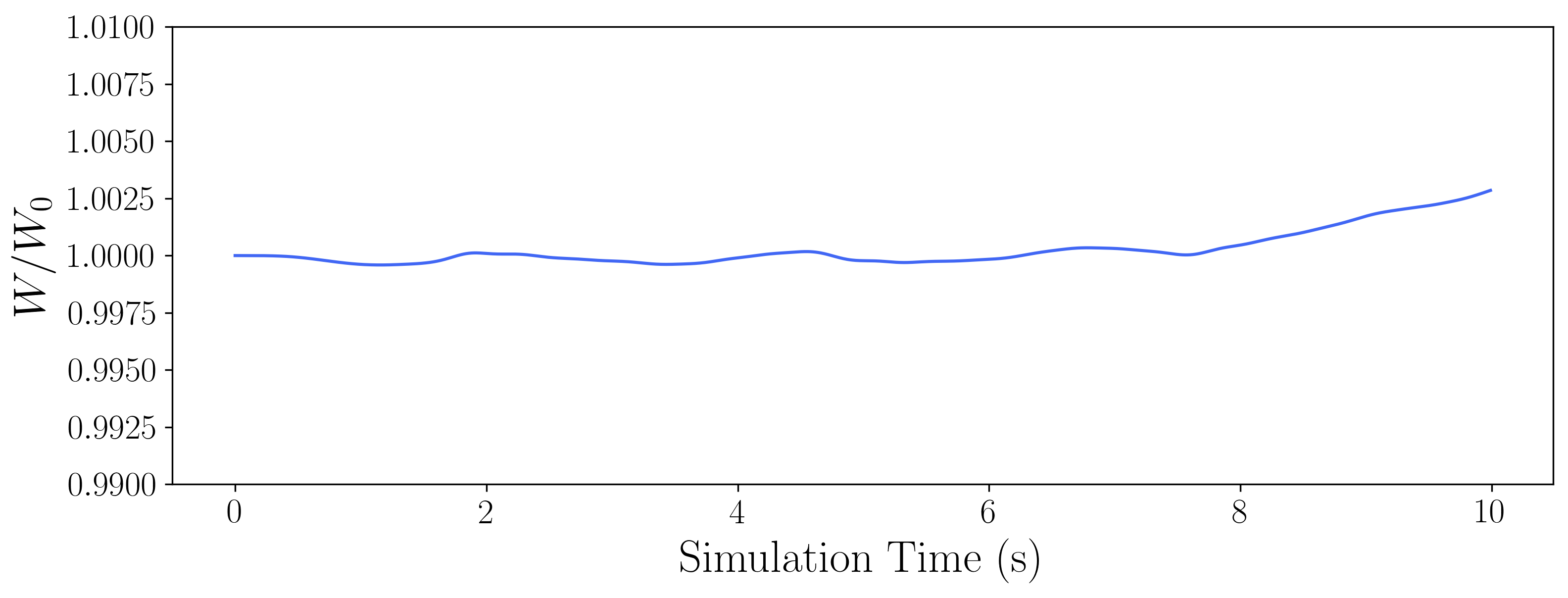}
    \caption{Evolution of the stellar internal energy for an $n=1$ polytrope under the equilibrium conditions. 
    Internal kinetic energy, $W$, is normalized by the initial condition for energy, $W_0$.}
    \label{fig:stellar_work}
\end{figure}

Throughout the simulation, we monitor the evolution of the internal kinetic energy, defined as 
\begin{equation}
W = \int_V PdV
\label{eq:stellar_Wrk}
\end{equation}
where the pressure $P$ is integrated over the stellar volume. Changes to the internal energy give indication of the degree to which the virial theorem is satisfied assuming a static gravitational potential and constant potential energy. We find initial perturbations present in the stellar pressure and density due to the relaxation of interpolated values in the discretized computational domain. These perturbations manifest as minute variations in the internal energy, and Figure \ref{fig:stellar_work} shows the evolution of the normalized internal energy. The amplitude of these changes to the internal energy are on the order of $0.25\%$ over $12 s$ and $1.2 \times 10^6$ computational timesteps. This small increase and energy does not impact large-scale stellar structure as the simulation is allowed to evolve. We conclude that our simulation is at 
equilibrium and supports the structural equations for an $n=1$ polytrope.  

\section{Magnetic Field Model}

\subsection{Hydromagnetic Conditions}
We follow the mixed-field solution of \citet{haskell_2008} and \citet{Roxburgh} which determine magnetic field solutions for spherically symmetric stars, treating the magnetic energy as a perturbation of the total stellar energy. \citet{haskell_2008} show that if one adds the ellipticity of a purely-poloidal field to the ellipticity introduced by a purely-toroidal field, the result will be different from the resulting ellipticity for a mixed magnetic field model. In this model, the equation of hydrostatic equilibrium is modified to include a magnetic term due to the Lorentz force as 
\begin{equation}
    \frac{\nabla P}{\rho}+\nabla\Phi_g =\frac{(\nabla\times\boldsymbol{B})\times \boldsymbol{B}}{4\pi\rho}= \frac{\boldsymbol{L}}{4\pi\rho},
    \label{eq:hydromagnetic_eq}
\end{equation}
 where $P$ is the pressure, $\rho$ is the density, $\Phi_g$ is the gravitational potential, $\boldsymbol{B}$ is the vector-valued magnetic field, and $\boldsymbol{L}$ is the Lorentz force. We refer to Equation \ref{eq:hydromagnetic_eq} as the equation for hydromagnetic equilibrium.

\citet{Roxburgh} shows that magnetic field configurations satisfying Equation \ref{eq:hydromagnetic_eq} must meet an additional constraint, determined by taking the curl of the equation for hydromagnetic equilibrium. Since the curl of a gradient is zero, i.e. $\nabla\times\nabla A = 0$, where $A$ is any scalar-valued function, the left-hand side of Equation \ref{eq:hydromagnetic_eq} vanishes under such operation, and we arrive at the following constraint:
\begin{equation}
    \nabla\times\left[\frac{\boldsymbol{B}\times(\nabla\times\boldsymbol{B})}{\rho}\right] = 0,
    \label{eq:mag_constraint}
\end{equation}
where $\rho$ is a barotropic EOS. Because the density $\rho$ is present in this constraint, Equation \ref{eq:mag_constraint} must be solved alongside Equation \ref{eq:hydromagnetic_eq}. Thus, the choice of barotropic EOS constrains allowable magnetic field models, motivating our reasoning for choosing an EOS and determining structural equations in the preceding section. Subsequent discussion will be limited to magnetic field solutions which adhere to the EOS determined for the $n=1$ polytrope in Equations \ref{eq:rho_eqn} and \ref{eq:prs_eqn}.


\subsection{Mixed Field Equations}
The choice of magnetic field configuration requires careful consideration of dynamically stable models which preserve field geometry under evolution and unallowable configurations which rapidly evolve and alter the stellar field structure. \cite{Flowers} discuss the instability of pure-poloidal stellar magnetic fields with uniform, unclosed field lines in the stellar interior. Pure toroidal configurations are also unstable, as instabilities form along the magnetic axis \citep{Tayler_toroidal_instability}. Mixed magnetic fields, including both poloidal and toroidal components, offer promising stable configurations (\cite{Flowers}, \cite{braithwaite_spruit_2006}, \cite{Bra09}, \cite{Yos19}). 

With consideration to stable field configurations, we adopt the axisymmetric mixed poloidal-toroidal magnetic field model of \citet{haskell_2008} for an $n=1$ polytropic EOS.  In this model, the magnetic field is divided into poloidal ($\boldsymbol{B_p}$) and toroidal ($\boldsymbol{B_t}$) components, where $\boldsymbol{B}_p = (B_{r}, B_{\theta}, 0$) and $\boldsymbol{B}_t = (0, 0, B_{\phi})$.  Magnetic field solutions take the form of 

\begin{equation}
    B_r = \frac{1}{r^2 sin\theta}\frac{\partial S}{\partial \theta}, 
    \label{eq:br}
\end{equation}
\begin{equation}
    B_\theta = -\frac{1}{r sin\theta}\frac{\partial S}{\partial r}, 
    \label{eq:btheta}
\end{equation}
\begin{equation}
    B_\phi = \frac{\beta(S)}{r sin\theta},  
    \label{eq:bphi}
\end{equation}
where $S(r,\theta)$ is a stream function connecting the poloidal and toroidal field components and $\beta$ is some function of the stream function.  Following \citet{Roxburgh}, we adopt 
\begin{equation}
    \beta = \frac{\pi \lambda}{R}S.
\end{equation}
We adopt a stream function of form 
\begin{equation}
    S(r,\theta) = A(r) sin^2\theta,
\end{equation}
where $A$ is 
\begin{multline}
    A = \frac{B_kR_{\star}^2}{(\lambda^2-1)^2y}\Big[2\pi\frac{\lambda y\cos{(\lambda y)}-\sin{(\lambda y)}}{\pi\lambda\cos{(\pi\lambda)}-\sin{(\pi\lambda)}} \\ +\left((1-\lambda^2)y^2-2\right)\sin{(y)} +2y\cos{(y)}\Big]. 
    \label{eq:A}
\end{multline}
In this expression, the constant $B_k$ sets the strength of the magnetic field, $R$ is the radius of the star, $y = \pi r/ R$ is a dimensionless radius, and $\lambda$ is an eigenvalue solution that sets the relative strengths of the poloidal and toroidal field components, where higher values for $\lambda$ correspond to a stronger toroidal component. We implement the first eigenvalue solution, $\lambda = 2.362$, as \citet{haskell_2008} show that ellipticity increases as the chosen eigenvalue increases. Thus, our implementation represents a limiting scenario for ellipticity resulting from the field expressions provided by \citet{haskell_2008}.  The field strength at the stellar surface, which we label $B_s$, imposes a constraint on the value of $B_k$, as we wish for the value of $B_s$ to adhere to magnetar surface field strengths of order $10^{15}$ G. By computing the average field strength for voxels along the surface of the stellar medium in the computational domain, we experimentally determine that a value of $B_k = 8\times 10^{16}$ G results in an average surface field strength $\bar{B}_s \sim 1.59\times 10^{15}$ G, with a maximum equal to $B_s^{\text{max}} \sim 2.02\times 10^{15}$ G in the equatorial plane of the star. This assignment is consistent with the notion that internal magnetic field strengths can range up to a few orders of magnitude higher than surface fields (\cite{Haensel, Bra09, Akg13,Yos19}).    

For the $n=1$ polytrope, dipolar solutions to the mixed magnetic field model take the form
\begin{equation}
    (B_r, B_\theta, B_\phi) = \left\{\frac{2A\cos{\theta}}{r^2}, \frac{-A'\sin{\theta}}{r}, \frac{\pi\lambda A\sin{\theta}}{rR_{\star}}\right\}.
    \label{eq:B_field_expression}
\end{equation}

It should be noted that for the barotropic EOS that we have adopted, boundary conditions require that all components of the mixed poloidal-toroidal field vanish at the surface of the star at $t=0$.  The interior mixed dipolar field can therefore only be matched to a vanishing external field.  We recognize that this is not physical, as fields are expected to be dipolar far from the star.  A large body of work has shown that MHD equilibria in stellar systems is greatly influenced by the choice of a barotropic vs a non-barotropic EOS (e.g., \cite{Rei09, Mas11, Mas15, gla16}).  These studies show that limitations on field configuration can be relaxed by the adoption of a non-barotropic EOS.  \citet{Mas11} perform a calculation of the ellipticity of an $n=1$ polytropic star that is not constrained to a barotropic EOS.  The authors compare their results to those of \citet{haskell_2008} for the same poloidal-to-total field ratios (but different field configurations) and find a difference of up to approximately one order of magnitude in the resulting ellipticity.  In general, they find larger values of $\epsilon$ for smaller $\lambda$ eigenvalues and smaller $\epsilon$ for larger $\lambda$.   
We acknowledge that our adoption of a barotropic EOS imposes a restriction on allowable field configurations and requires that the poloidal and toroidal components of the field are governed by a single equation.  While a non-barotropic EOS would allow these parameters to be defined independently, the eigenvalue $\lambda$ allows for the relative strengths of these components to be specified.  Although the eigenvalue solutions are discrete, we still retain the ability to find solutions with varying poloidal and toroidal field components.  We therefore believe that, although simplistic, our choice of EOS is appropriate as an initial step toward developing a more complex model in future investigations.  For a full discussion of boundary conditions and restrictions on the field configuration in our framework, see \citet{haskell_2008}.   


\subsection{Magnetic Field Model Stability Validation}
The stability of a magnetic field configuration is dependent on its evolution under Alfv\'en time scales, which define the period necessary for tension-induced Alfv\'en waves to propagate throughout the magnetic field. These waves determine the geometric evolution of the field configuration, and thus provide a strong basis for studying the stability of stellar magnetic fields \citep{Goedbloed}. For a homogeneous plasma with uniform density $\rho_0$ and magnetic field strength $B_0$, the velocity of an Alfv\'en wave is
\begin{equation}
    v_A = \frac{B_0}{\sqrt{\mu_0\rho_0}}.
    \label{eq:alfven_vel}
\end{equation}
As density and magnetic field strength vary in our model, we determine a volume averaged value for the Alfv\'en velocity, $\bar{v}_A$, where $\bar{v}_A \approx 2.736\times 10^9$ cm$\cdot$s$^{-1}$. The Alfv\'en crossing time for wave propagation is then
\begin{equation}
    t_A = \frac{d}{v_A},
\end{equation}
where $d$ is the wavelength of the Alfv\'en wave, which is approximated in the stellar interior by the radius, $ R_{\star} = 10$ km \citep{suzuki_nagataki}. We compute the volume-averaged Alfv\'en crossing time for our model to be $\bar{t}_A\approx 0.4$ ms, in agreement with prior evaluation of the Alfv\'en crossing time for interior magnetic fields in highly magnetic neutron stars \citep{suzuki_nagataki}. 

In assessing the stability of our model's magnetic field configuration, the computed Alfv\'en crossing time indicates that robust analysis of the field's stability may be conducted by analyzing the field configuration after several Alfv\'en crossings. We conduct stability analysis of the magnetic field configuration by comparing the geometry of the initial field configuration to the evolved state after 100 Alfv\'en crossings. 

\begin{figure}
    \includegraphics[width=\columnwidth]{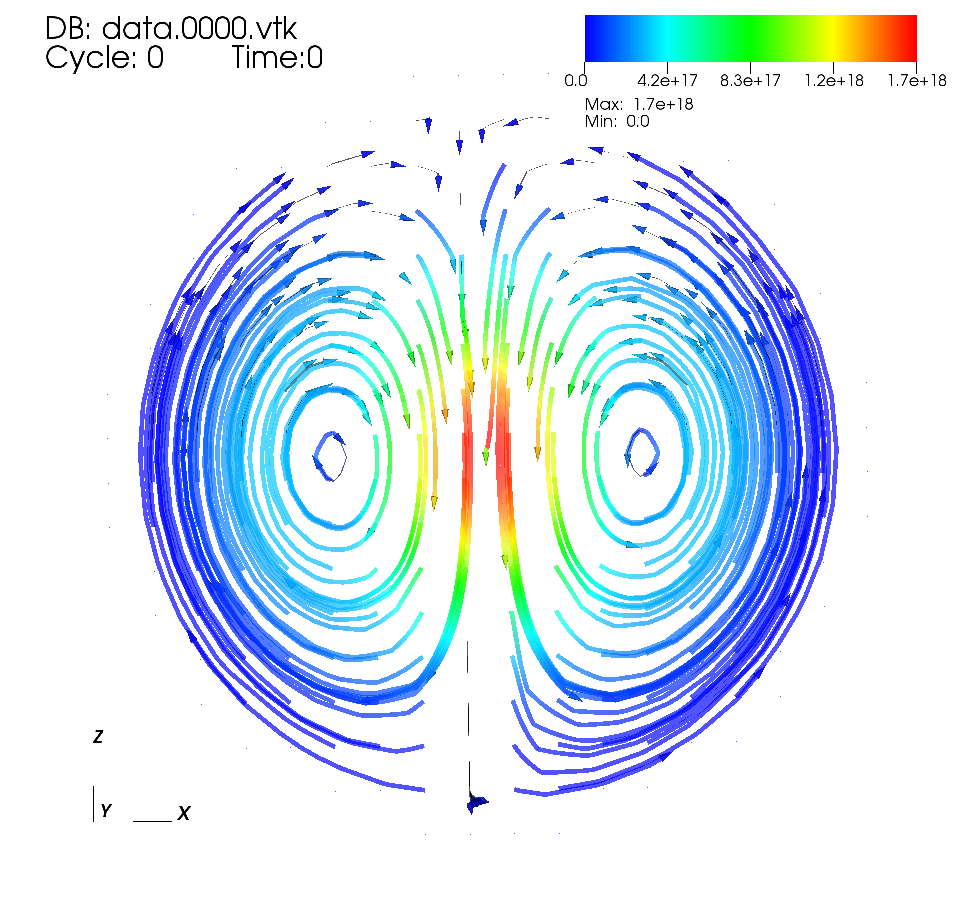}
    \caption{Streamlines for the initial configuration of the poloidal component field. The colormap corresponds to the strength of the magnetic field in gauss along computed streamlines.} 
    \label{fig:stream_pol_0}
\end{figure}

\begin{figure}
    \includegraphics[width=\columnwidth]{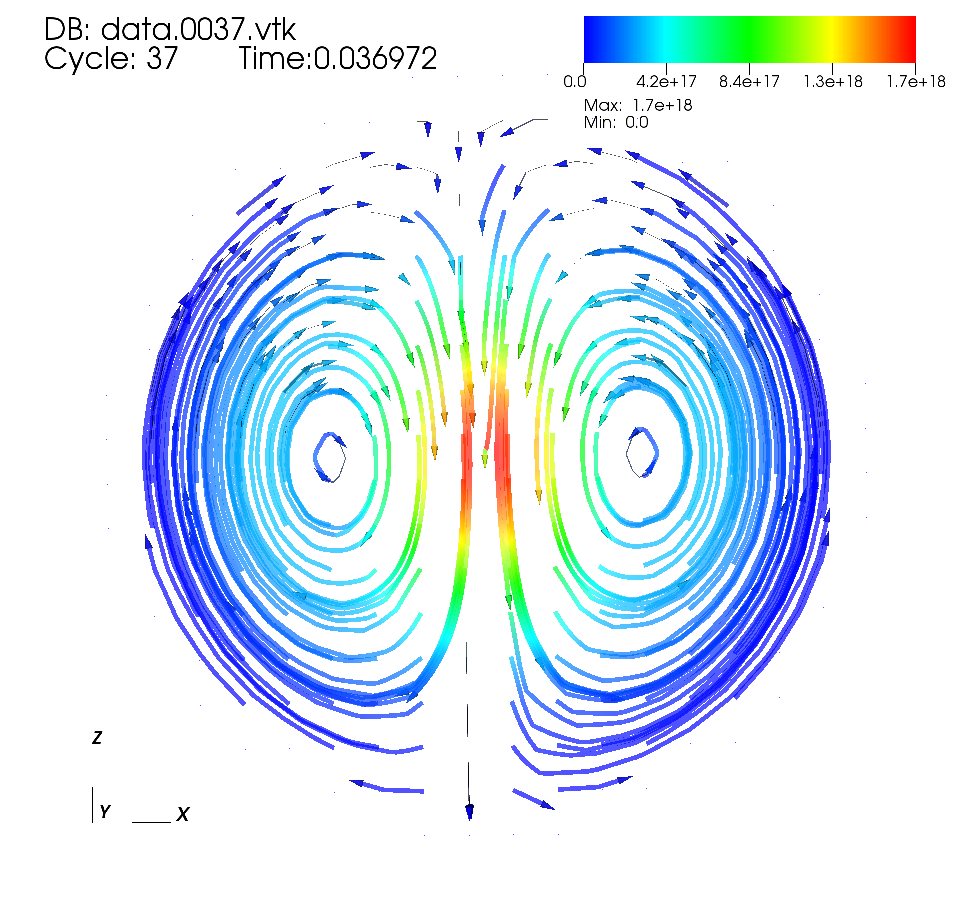}
    \caption{Streamlines for the poloidal component field after 100 Alfv\'en crossings.  Field strengths are given in gauss.}
    \label{fig:stream_pol_37}
\end{figure}

\begin{figure}
    \includegraphics[width=\columnwidth]{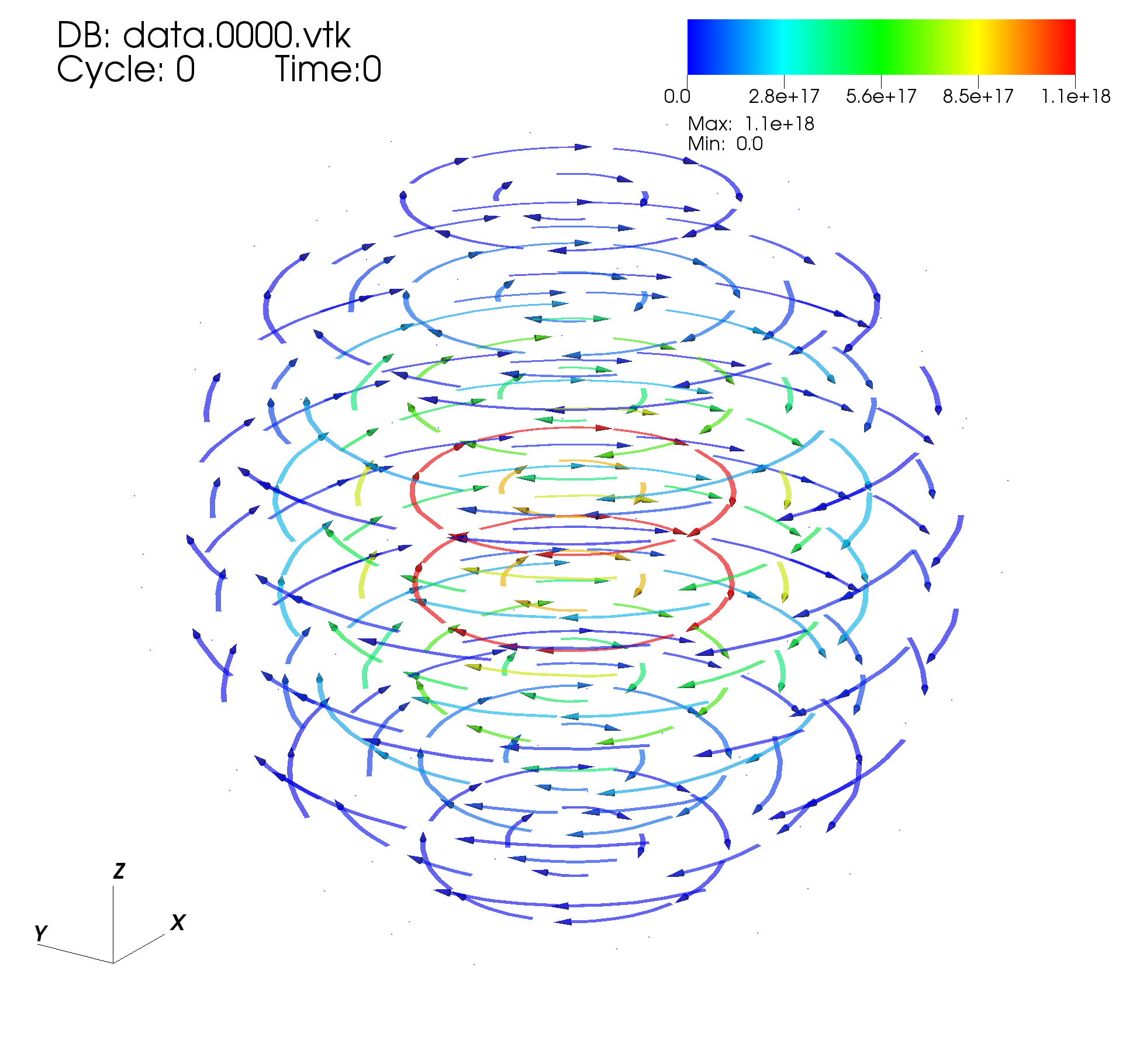}
    \caption{Streamlines for the initial configuration of the toroidal component field. The colormap corresponds to the strength of the magnetic field in gauss along computed streamlines. }
    \label{fig:stream_tor_0}
\end{figure}

\begin{figure}
    \includegraphics[width=\columnwidth]{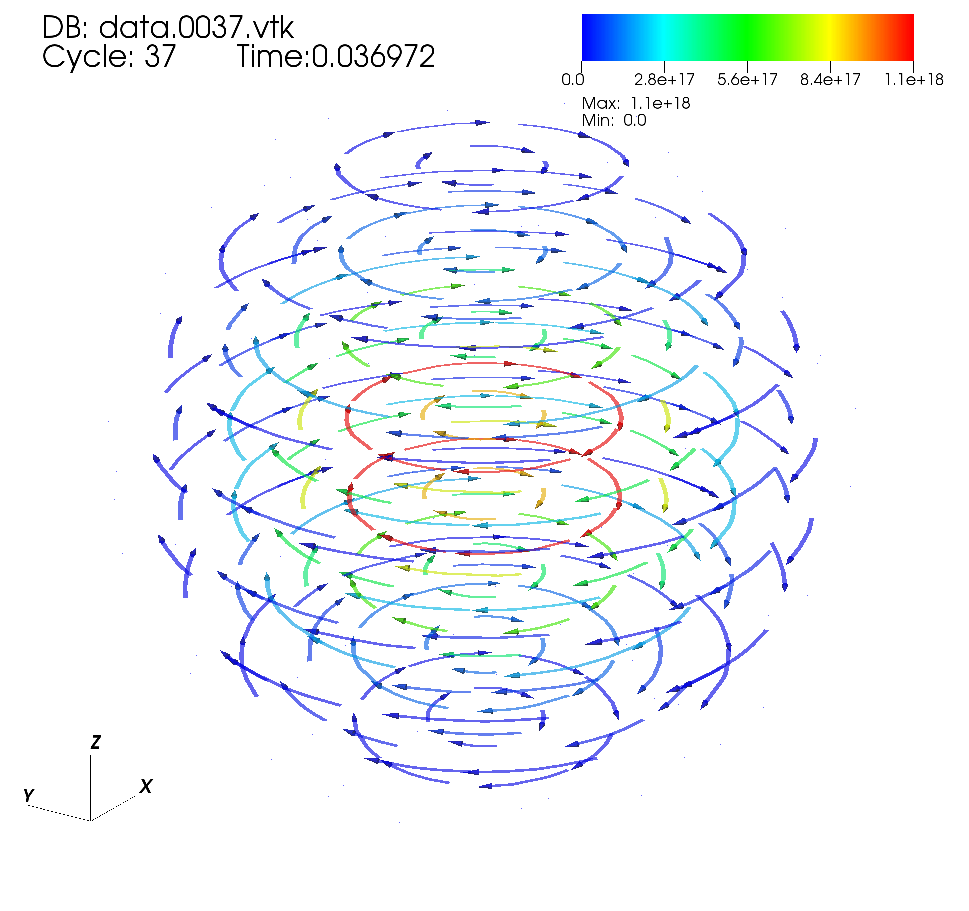}
    \caption{Streamlines for the toroidal component field after 100 Alfv\'en crossings.  Field strengths are given in gauss. }
    \label{fig:stream_tor_37}
\end{figure}

We use streamlines to label the geometry of the field. Streamlines represent the trajectories of fluid elements in the presence of an axisymmetric stellar magnetic field, and evolution of their form provides immediate awareness of changes to the magnetic field structure. We plot streamlines for both the poloidal component field in Figures \ref{fig:stream_pol_0} and \ref{fig:stream_pol_37} and for the toroidal component field in Figures \ref{fig:stream_tor_0} and \ref{fig:stream_tor_37}. After 100 Alfv\'en crossings, consistent arrangement of magnetic field streamlines indicate little change in the structure of the field, and we conclude that the field configuration is well preserved. Our findings provide evidence that the chosen magnetic field equations (\ref{eq:B_field_expression} and \ref{eq:A}) correspond to a stable configuration. 




\section{Determination of Magnetar Ellipticity}
A star with principal moment of inertia $I_0$ about its axis of symmetry will produce gravitational radiation if the axis of rotation is offset from the symmetry axis. The symmetry axis will freely precess about the rotational axis, and gravitational radiation will be produced with wave strain 
  \begin{equation}
      h_0 = \frac{4\pi^2G}{c^4}\frac{I_{0}f^2_{\text{gw}}}{r}\epsilon,
      \label{eq:wavestrain}
  \end{equation}
  where $c$ is the speed of light and $r$ is the stellar distance.  The quantity $f_{\text{gw}}$ is the gravitational wave frequency, equal to twice the rotational frequency of the star. Here, we assume that the rotational axis, taken to be the $z$-axis, is optimally pointed towards an observer on Earth.  The ellipticity, $\epsilon$, is a measure of stellar deformation and is defined as 
  \begin{equation}
     \epsilon = \frac{I_{zz} - I_{xx}}{I_0}. 
     \label{eq:ellip}
  \end{equation}
$I_0$ is the moment of inertia of the unperturbed spherical star, and $I_{zz}$ and $I_{xx}$ are principal moments of inertia, determined via the inertia tensor, 
  \begin{equation}
      I_{jk} = \int_V \rho(r)\big(r^2\delta_{jk}-x_jx_k\big)dV.
      \label{eq:inertiatens}
  \end{equation}
 If the stellar ellipticity is negative, whereby $I_{zz} < I_{xx}$, the star is considered \textit{prolate}. Conversely, a positive stellar ellipticity, such that $I_{zz} > I_{xx}$, corresponds to an \textit{oblate} star.

    Calculation of the continuous GW strain, $h_0$, depends on the degree to which the distribution of mass is spherically non-uniform about the rotational axis, i.e, when $I_{zz}\neq I_{xx}$. The presence of a strong internal magnetic field in magnetars modifies the equilibrium configuration of the stellar structure, perturbing the density profile, $\rho$, through quadrupolar ($\ell=2$) deformations. Perturbation of the density changes the ellipticity via modification of the principal moments of inertia: $I_{xx}$, $I_{yy}$, and $I_{zz}$.
    
  \subsection{Modeling Deformations in the Computational Domain }
  We use the astrophysical fluid dynamics code \texttt{PLUTO} \citep{PLUTO} to specify and simulate our computational stellar model. Flux computation is made via the Hartman, Lax, Van Leer (\texttt{hllc}) solver. The data visualization platform \texttt{VisIt} \citep{HPV:VisIt} is used to analyze simulation data, including evaluation of the moment of inertia tensor in a specified computational domain. We determine $I_{zz}$ and $I_{xx}$ by evaluating the moment of inertia tensor over domain voxels for which $r < R_{\star}$ at simulation time steps of 100 ms. 
  
As the inertia tensor (Equation \ref{eq:inertiatens}) is a volume integral over the stellar interior, we anticipate numerical limitations on the accuracy of computed values for $I_{zz}$ and $I_{xx}$, as each tensor component must be computed over a discretized  domain of finite three-dimensional voxels. 
  
  For the initial configuration of the stellar model at $t = 0$ s, the inertia tensor is expressed as 
    \begin{equation}
          I_{jk} = \int_V \rho(r,t=0)\big(r^2\delta_{jk}-x_jx_k\big)dV,
          \label{eq:Inaught_integral}
    \end{equation}
    where $\rho(r,t=0)$ takes the form of Equation \ref{eq:rho_eqn}, such that the analytic evaluation of Equation \ref{eq:Inaught_integral} for all principal moments of inertia gives
    \begin{equation}
        I_0  = I_{xx} = I_{yy} = I_{zz} = \frac{8(\pi^2-6)R_{\star}^5\rho_c}{3\pi^3}.
        \label{eq:Inaught_analytic_exp}
    \end{equation}

\begin{figure}
    \centering
    \includegraphics[width=\columnwidth]{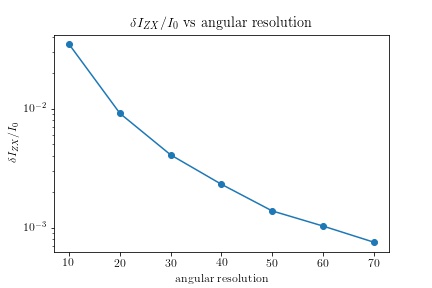}
    \caption{Estimation of the error arising from the discretization of the spatial region of computation. Angular resolution is given in number of divisions along the azimuthal plane. At $t = 0$, the conditions are set for a spherical star. The difference in moments of inertia $\delta I_{ZX} = |I_{ZZ} - I_{XX}|$ is scaled by the analytic moment of inertia $I_0$, therefore indicating the error contribution from discretization.}
    \label{fig:ang_error}
\end{figure}
Although higher angular resolution allows greater precision in both $I_{xx}$ and $I_{zz}$, such improvements come at the cost of greater wall time, or the elapsed real time necessary to complete a computational modelling run through a specified simulation duration. Thus, consideration is given to balancing the trade-off between resolution and compute time, and an angular resolution of $n_{\theta,\phi} = 30$ is selected for simulations. 

\begin{figure}
    \centering
    \includegraphics[width=\columnwidth]{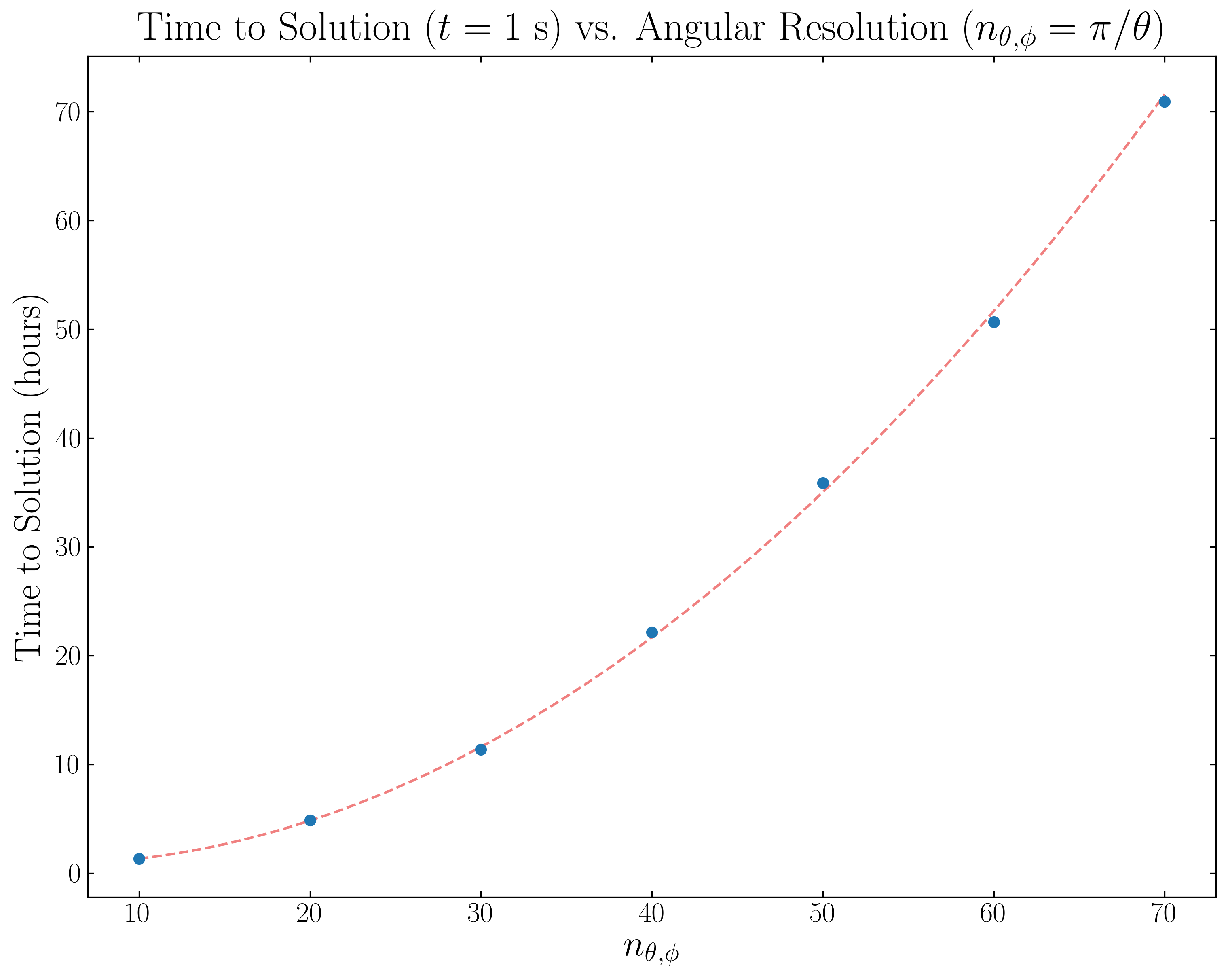}
    \caption{Quadratic regression of elapsed real (wall) time to solution ($t = 1$ s) against angular resolution for a single core computation (Intel(R) Xeon(R) CPU E5-2640 v3 @ 2.60GHz). Higher resolution causes increasingly prohibitive compute time.}
    \label{fig:time_to_sol}
\end{figure}


Crucial to the evaluation of the ellipticity, $\epsilon$, is the difference $I_{zz} - I_{xx}$, which we refer to as $\Delta I$. Because the finite-difference integration scheme for these inertia tensor components over the spherical mesh provide slightly different values for $I_{zz}(t = 0)$ and $I_{xx}(t = 0)$, the value 
\begin{equation}
    \delta I_{zx} = |I_{zz}(t = 0)-I_{xx}(t = 0)|\neq 0
\end{equation} 
is of significance, representing a systematic error in our evaluation of $I_{xx}$ and $I_{zz}$. Therefore, we represent numerical evaluations for these tensor components as
\begin{equation}
     I_{xx} = I_{xx} \pm \delta I_{zx}
     \label{eq:Ixx_err_exprs}
\end{equation}
and
\begin{equation}
    I_{zz} = I_{zz} \pm \delta I_{zx}.
    \label{eq:Izz_err_exprs}
\end{equation}

The discretization error in numerically computing 
moments of inertia is determined via the difference between each numerically computed tensor component at simulation time $t = 0$ and the analytic result of Equation \ref{eq:Inaught_analytic_exp}. 
We plot the difference in moments of inertia $\delta I_{ZX}$ scaled by $I_0$, the numerical moment of inertia at $t = 0$, as a function of voxel resolution in Figure~\ref{fig:ang_error} .
The radial resolution of the spherical mesh is kept constant in each plot while the angular resolution, measured by the number of discretizations along the polar ($\theta$) and azimuthal ($\phi$) axes, expressed as  
\begin{equation}
    n_{\theta,\phi} = \frac{\pi}{d\theta} = \frac{2\pi}{d\phi},
    \label{eq:theta_res}
\end{equation}
varies from 10 to 70. We notice that the absolute error is proportional to $1/n_{\phi,\theta}^2$, with improvements to angular resolution affording increasingly less reduction in error for higher values of $n_{\phi,\theta}$. 
As the angular resolution increases, this error decreases. For improved results, a higher resolution is needed, which significantly increases the computation time, as shown in Figure~\ref{fig:time_to_sol}.

\subsection{Deformation Results} \label{sec:res_effect}
In order to determine whether our simulated deformation results are in accordance with expectation, we simulate the instance of stellar hydrostatic equilibrium by removing the magnetic field model. For the instance of hydrostatic equilibrium, the null hypothesis is that $I_{xx}$ and $I_{zz}$ do not change from their initial configuration, such that $I_{xx} = I_{zz}$ and $\epsilon  = 0$. 

\begin{figure}
    \centering
    \includegraphics[width=\columnwidth]{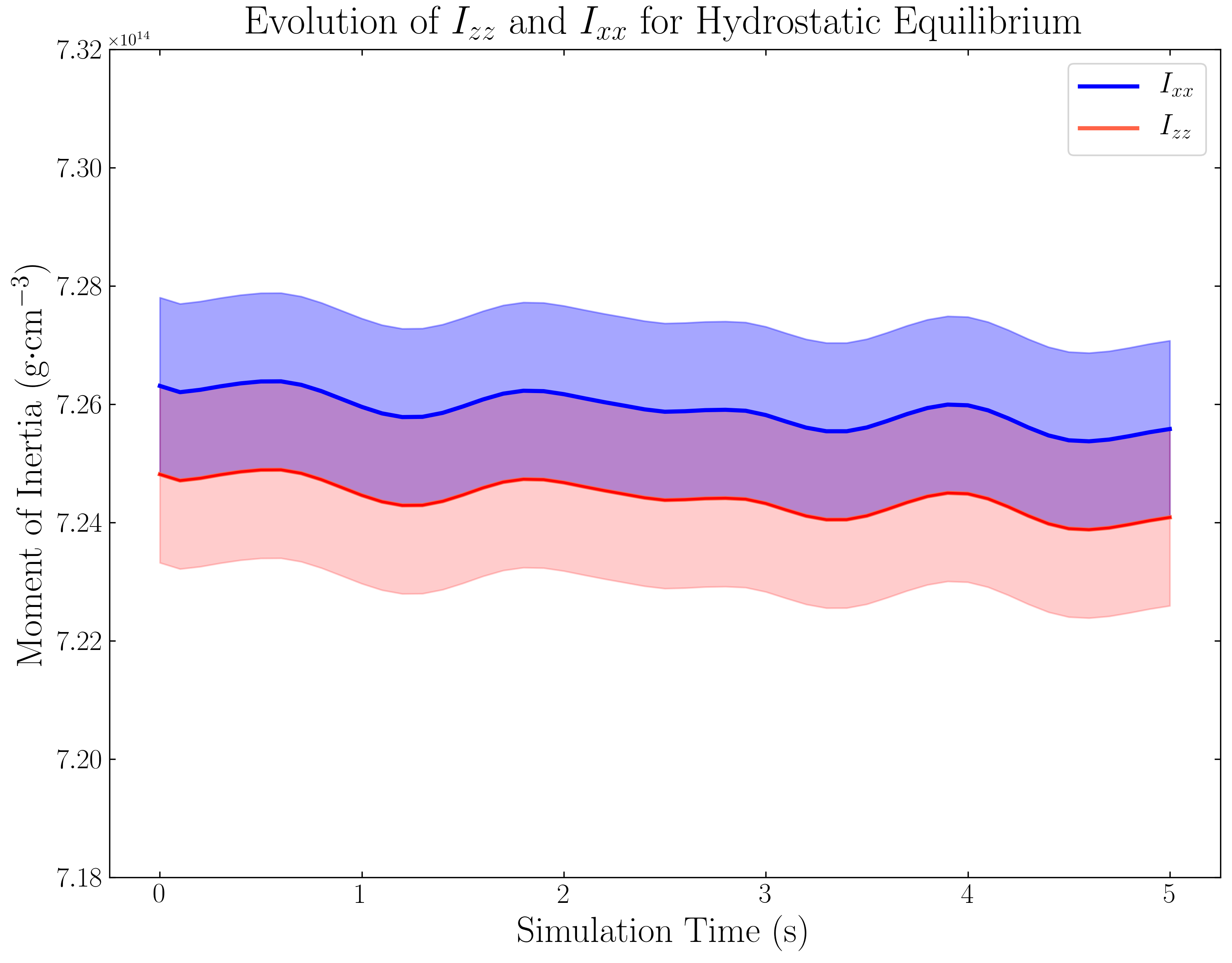}
    \caption{$I_{zz}$ and $I_{xx}$ evolution for conditions of hydrostatic equilibrium. The error margin $\delta I_{zx}$ is represented by the shaded regions of each curve. The precise overlap of each error margin suggests that fluctuations in the inertia tensor are perfectly symmetric under hydrostatic equilibrium, and that the ellipticity remains zero throughout the duration of the simulation.}
    \label{fig:hstatic_MOI}
\end{figure}

The evolution over simulation time of $I_{xx}$ and $I_{zz}$ for the instance of hydrostatic equilibrium is displayed in Figure \ref{fig:hstatic_MOI}. Both inertia components are assigned an error margin as expressed in equations \ref{eq:Ixx_err_exprs} and \ref{eq:Izz_err_exprs}, represented by the lighter shaded regions surrounding each curve. 

\begin{figure}
    \centering
    \includegraphics[width=\columnwidth]{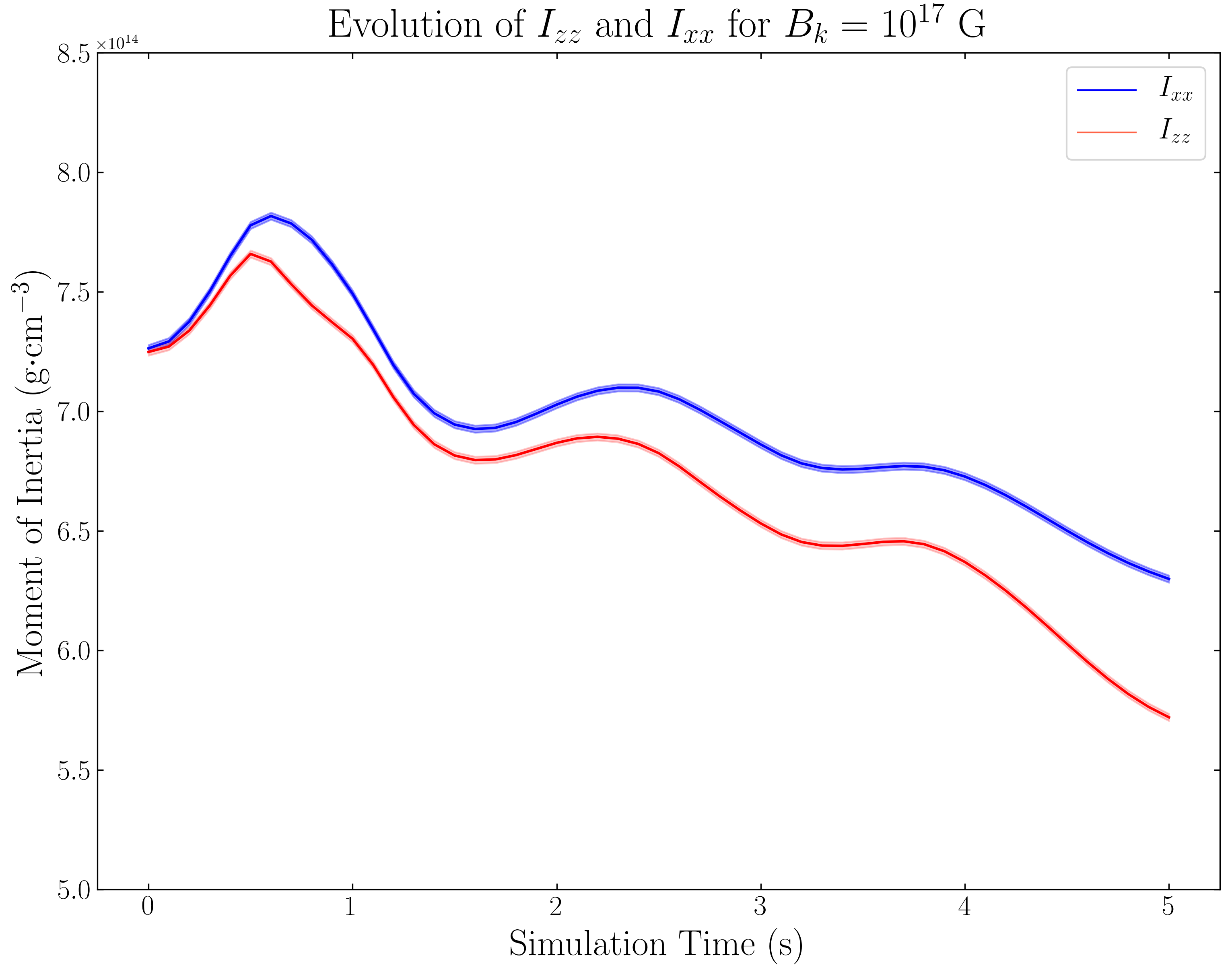}
    \caption{$I_{xx}$ and $I_{zz}$ evolution in the presence of a strong magnetic field. Notice that the inertia tensor components become distinctly separated, allowing measurement of ellipticity}
    \label{fig:MOI_b1e17}
\end{figure}

Because the error margins for both $I_{xx}$ and $I_{zz}$ overlap for the duration of the simulation, we strictly can not distinguish a non-zero value for the ellipticity, $\epsilon$. Thus, we verify the trivial null hypothesis for hydrostatic equilibrium. As an aside, we note that for Figures \ref{fig:hstatic_MOI} and \ref{fig:MOI_b1e17}, the moment of inertia is given in g$\cdot$cm$^{-3}$ because the choice of normalized radius, $R_{\star} = 1.0$, leaves equations for the moment of inertia such as the analytic result of Equation \ref{eq:Inaught_analytic_exp} with dimensions of density. 

For the non-trivial instance in which $I_{xx}$ and $I_{zz}$ evolve such that their error margins do not overlap, we can determine experimental measurements for ellipticity. We calculate error bounds for the ellipticity, where  
\begin{align}
\begin{split}
    \Delta I_{max} &= (I_{zz} + \delta I_{zx}) - (I_{xx} - \delta I_{zx}) \\
   \Delta I_{min} &= (I_{zz} - \delta I_{zx}) - I_{xx} + \delta I_{zx}),
\end{split}
\end{align}
such that an experimentally determined ellipticity, $\epsilon$, is expressed as  
\begin{equation}
    \epsilon = \frac{\Delta I}{I_0} \pm \frac{2\delta I_{zx}}{I_0}.
    \label{eq:exp_ellip}
\end{equation}

We reintroduce the magnetic field model and graph the evolution of $I_{xx}$ and $I_{zz}$ through a simulation time of $t = 5.0$ s in Figure \ref{fig:MOI_b1e17}, where the magnetic field is assigned the magnitude $B_k = 1\times10^{17}$ G in accordance with a surface field strength of order $10^{15}$ G. In distinct difference to the instance of hydrostatic equilibrium in Figure \ref{fig:hstatic_MOI}, $I_{xx}$ and $I_{zz}$ become distinguishable such that measurements of ellipticity can be performed.

\begin{figure}
    \centering
    \includegraphics[width=\columnwidth]{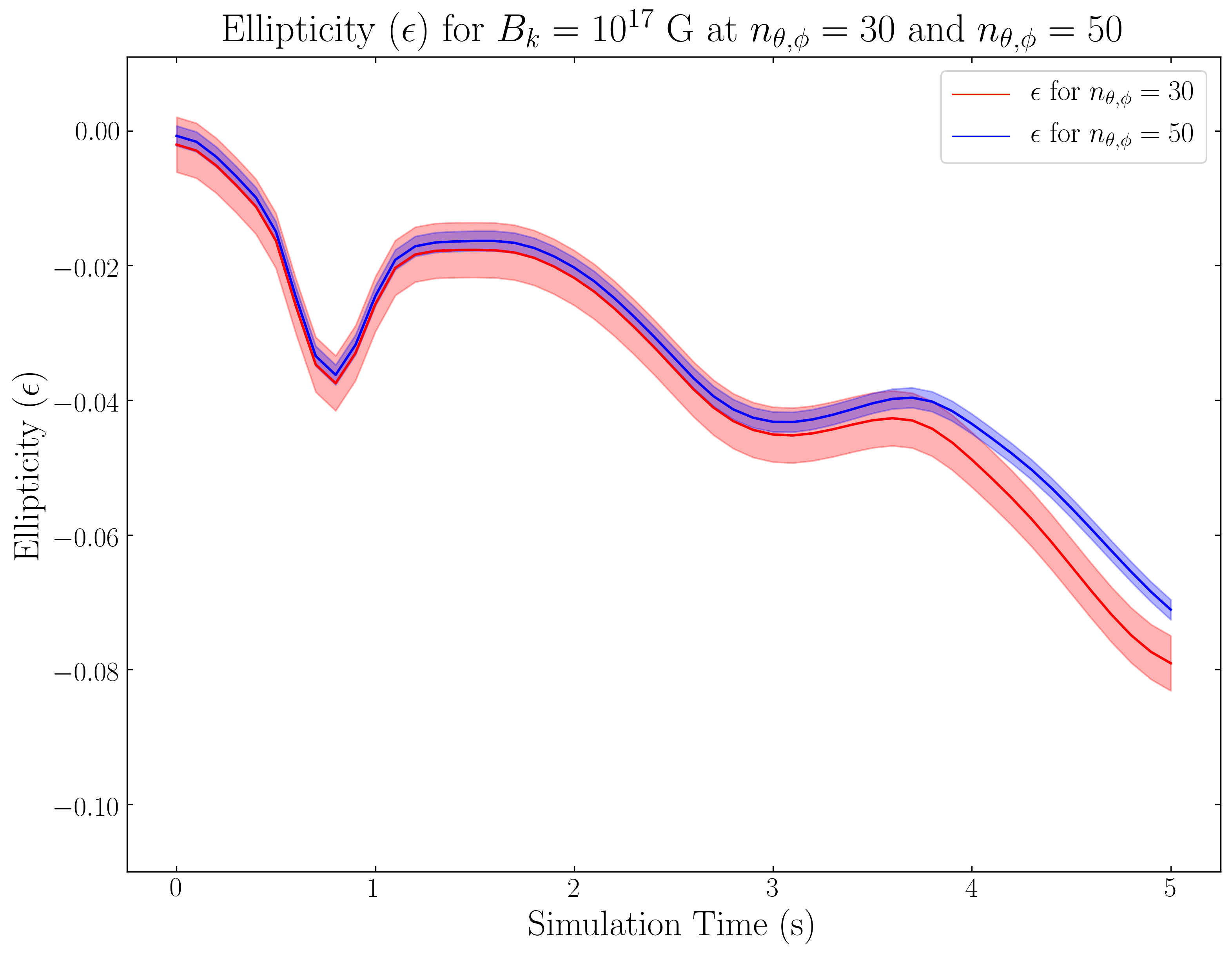}
    \caption{Comparison of ellipticity for angular resolution $n_{\theta,\phi} = 30$ (red) and  $n_{\theta,\phi} = 50$ (blue).}
    \label{fig:ellip_res_compar}
    \end{figure}

The computed ellipticity for $I_{xx}$ and $I_{yy}$ as shown in Figure \ref{fig:MOI_b1e17} is plotted in Figure \ref{fig:ellip_res_compar} for discretization resolutions of $n_{\theta,\phi} = 30$ and $n_{\theta,\phi} = 50$.  We find that the value of ellipticity becomes increasingly negative as the star becomes more prolate under evolution.  This result of a prolate star is in agreement with similar studies (e.g., \cite{haskell_2008, Mas11}).  For $t \lesssim 3$ s, our results agree for both higher and lower resolution ellipticity measurements within error margins set by the ellipticity error margin expressed in Equation \ref{eq:exp_ellip}.  For $t \gtrsim 3$ s, the ellipticity measurement for higher angular resolution data trends marginally less negative than the lower resolution counterpart.  This result suggests that increased angular resolution may result in values for the ellipticity that are closer to zero.  We find that for $n_{\theta,\phi} = 50$, the magnitude of ellipticity is $\epsilon(t = 5.0)\approx (7.11\pm 0.14) \times 10^{-2}$, while for $n_{\theta,\phi} = 30$, the ellipticity magnitude is $\epsilon(t = 5.0)\approx (7.91 \pm 0.41) \times 10^{-2}$.

 \subsection{Extended Simulation Results}
 
\begin{figure}
        \includegraphics[width=\columnwidth]{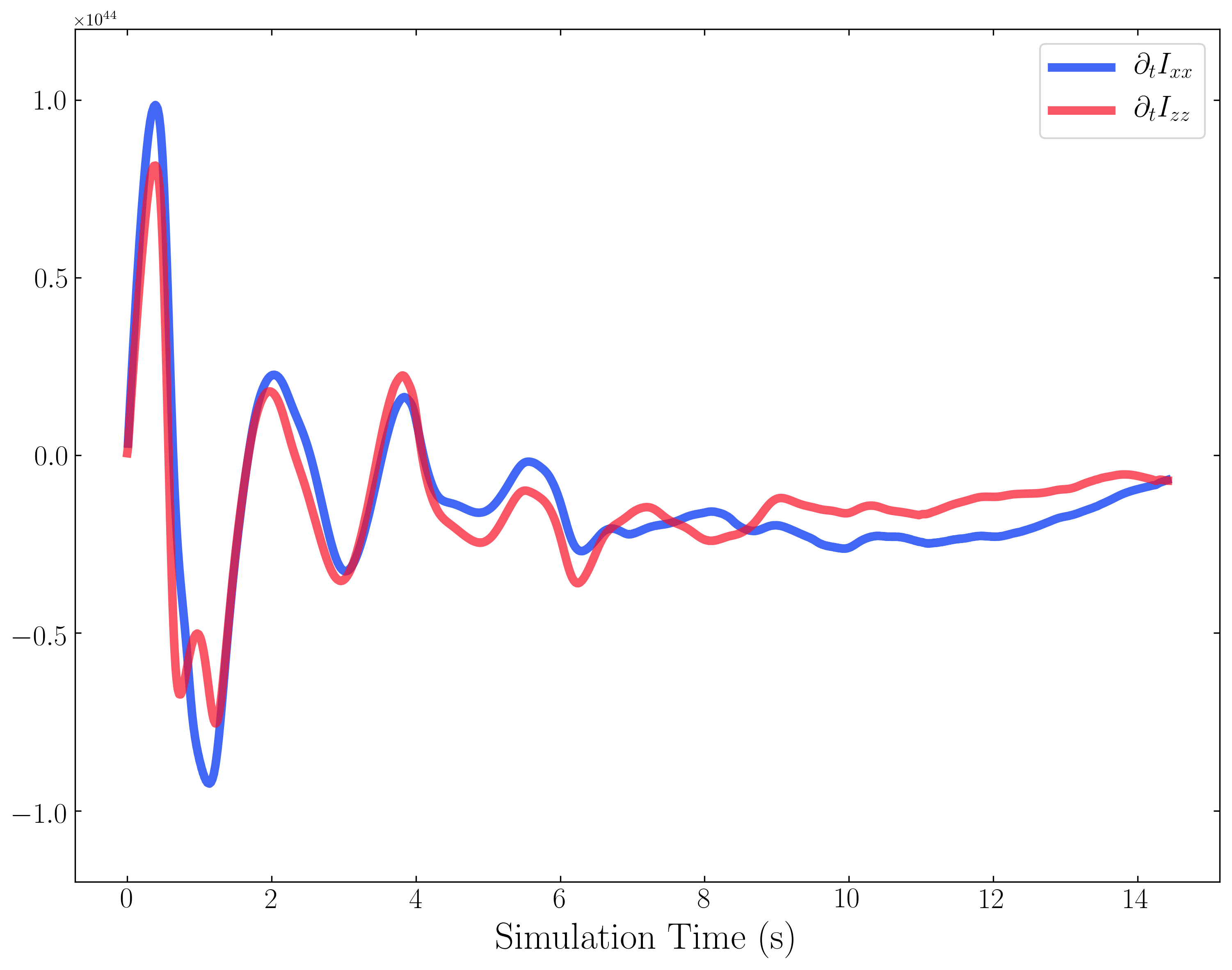}
        \caption{The first time derivative for $I_{xx}$ and $I_{zz}$ in an extended simulation through simulation time $t=14$ s. Dramatic evolution from the initial configuration the stellar medium is constrained as the star approaches MHD equilibrium.}
        \label{fig:MOI_first_deriv}
\end{figure}

\begin{figure}
        \includegraphics[width=\columnwidth]{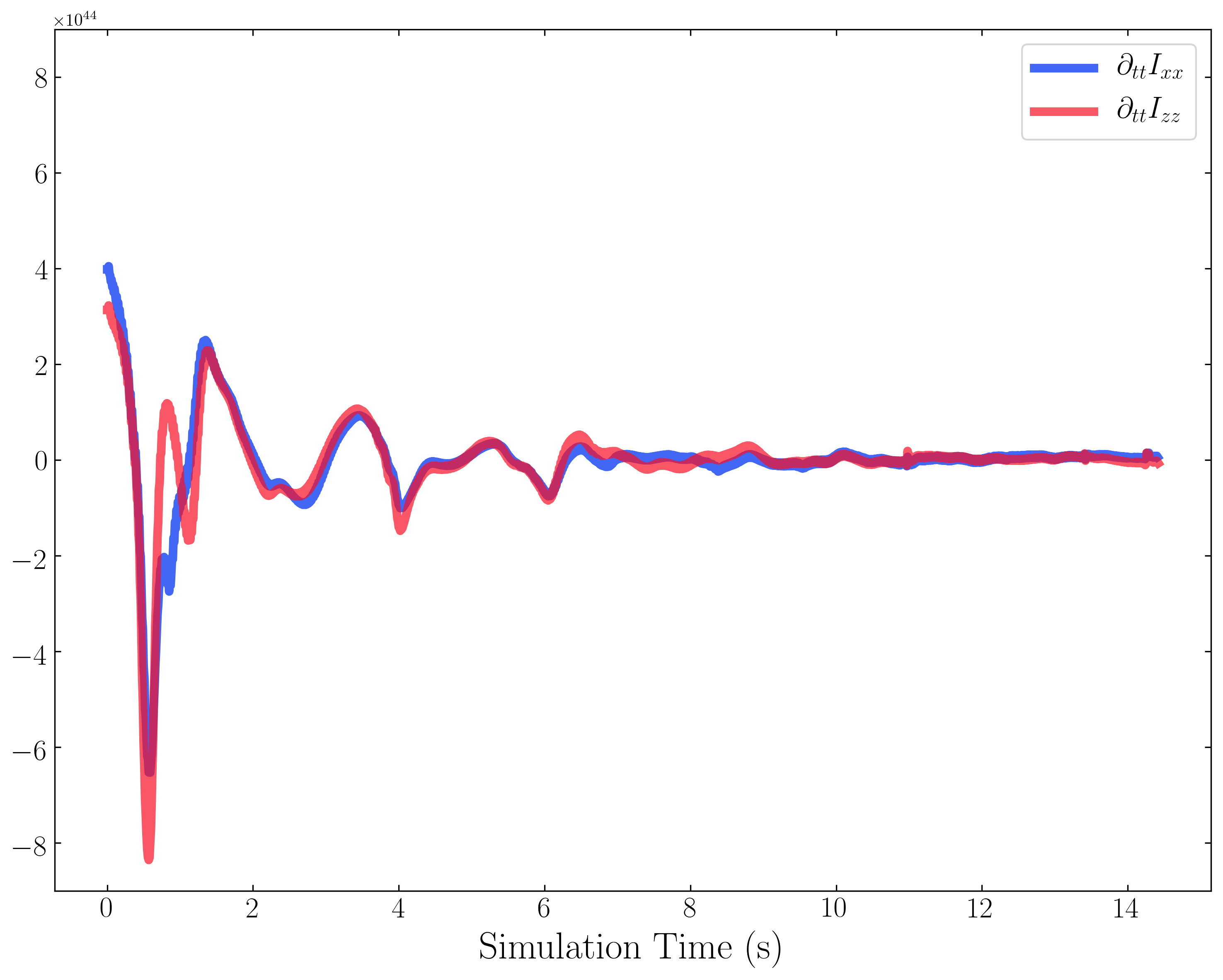}
        \caption{The second time derivative for $I_{xx}$ and $I_{zz}$ in an extended simulation through simulation time $t=14$ s. The rapid decrease in the magnitude of each time derivative appears highly constrained by a decaying envelope as the star approaches MHD equilibrium.}
        \label{fig:MOI_second_deriv}
\end{figure}

 The continual evolution of $I_{xx}$ and $I_{zz}$ through simulation time $t=5$ s motivates us to run an extended simulation through time $t = 14$ s. We analyze the first and second derivative of $I_{xx}$ and $I_{zz}$ to determine whether extended simulation time indicates that the evolution of these principal moments of inertia are constrained as the star approaches MHD equilibrium. In Figures \ref{fig:MOI_first_deriv} and \ref{fig:MOI_second_deriv}, we plot the first and second time derivative for $I_{xx}$ and $I_{zz}$, where we find strong evidence of a decaying envelope which constrains the evolution of each moment of inertia.
 
 Equation \ref{eq:ellip} relates the ellipticity to the principal moments of inertia $I_{zz}$ and $I_{xx}$. Taking a derivative of the equation for ellipticity with respect to time, we trivially find that  
 
 \begin{equation}
    \frac{\partial \epsilon}{\partial t} \propto \frac{\partial I_{zz}}{\partial t} -  \frac{\partial I_{xx}}{\partial t}. 
     \label{eq:eps_deriv}
 \end{equation}
 
 As the time derivative of stellar ellipticity is proportional to the difference between the time derivatives of the principal moments of inertia,
 a stellar medium which approaches MHD equilibrium (whereby $\partial _t(I_{zz})$, $\partial_t (I_{xx})$, and higher order derivatives approach zero) will also approach constant ellipticity. 
 
 Our results for the evolution of the first and second time derivatives for $I_{xx}$ and $I_{zz}$ indicate that the timescale for perturbation of the stellar structure to be strongly damped by MHD forces is of order 10 s.  
 
\begin{figure}
        \includegraphics[width=\columnwidth]{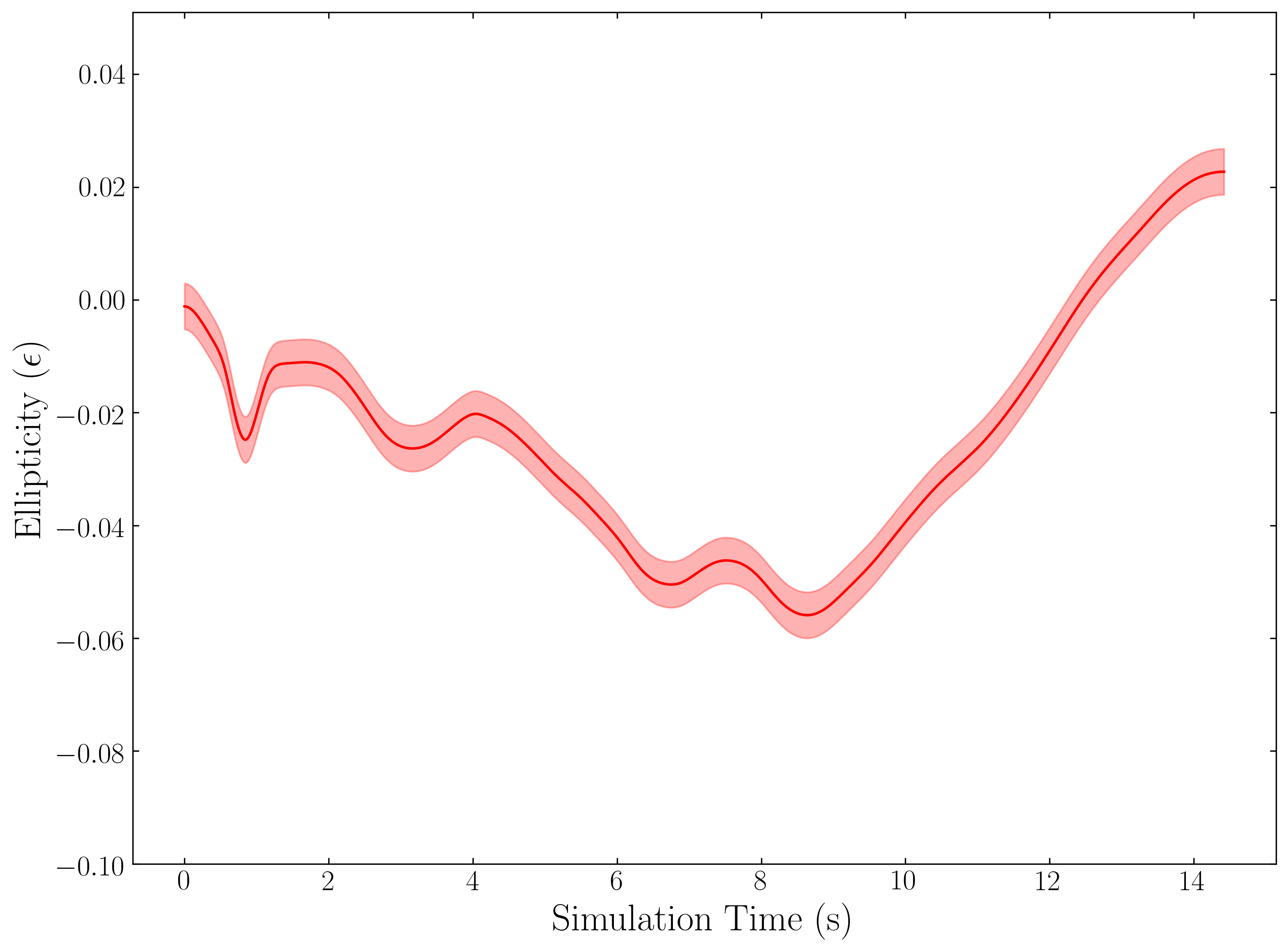}
        \caption{Stellar ellipticity in an extended simulation through simulation time $t = 14$ s.}
        \label{fig:eps_14s}
\end{figure}

 We compute the stellar ellipticity for our extended simulation and plot our results in Figure \ref{fig:eps_14s}. We find that over the course of our simulation, the maximum magnitude of the ellipticity is $\sim5.6\times 10^{-2}$. While the ellipticity continues to evolve dynamically over the course of the simulation, analysis of the evolution of the principal moments of inertia $I_{xx}$ and $I_{zz}$ provide strong evidence of stabilization and future evolution of the ellipticity will be constrained as the stellar medium nears MHD equilibrium. 
 
 

\subsection{Upper-Limit Estimates for Magnetar Gravitational Wave Strain}

With the adoption of a canonical value for the unperturbed stellar moment of inertia, $I_0$, the gravitational wave strain (Equation \ref{eq:wavestrain}) can be calculated with knowledge of three stellar parameters: ellipticity, rotational period, and distance to the source.  Our result for ellipticity is combined with data for rotational period and distance for individual magnetars to calculate upper limits to the wave strain for sources in the McGill Magnetar Catalog \citep{mcgill}.  

The ellipticity results presented in this paper are determined for a magnetar with surface field strength $B_s \approx 2.0\times 10^{15}$ G.  Our findings in Section \ref{sec:res_effect} lead us to adopt $\epsilon\approx 7.11\times 10^{-2}$ under the assumption that higher resolution simulation results in a more precise determination of ellipticity. Following the work of \citet{lasky_2015}, the wave strain can be calculated via 
\begin{equation}
    |h_0| = 4.2\times10^{-26}\left(\frac{\epsilon}{10^{-6}}\right)\left(\frac{\tau}{10 \text{ ms}}\right)^{-2}\left(\frac{d}{1 \text{ kpc}}\right)^{-1},
    \label{eq:wavestrain_exp}
\end{equation}
with rotational period in units of ms and distance in kpc.  Our computed wave strain estimates are listed in Table \ref{tab:mcgill_sources} with the exception of catalog source MG J1833-0831 due to the lack of data for the source's stellar distance.  

\begin{figure}
        \includegraphics[width=\columnwidth]{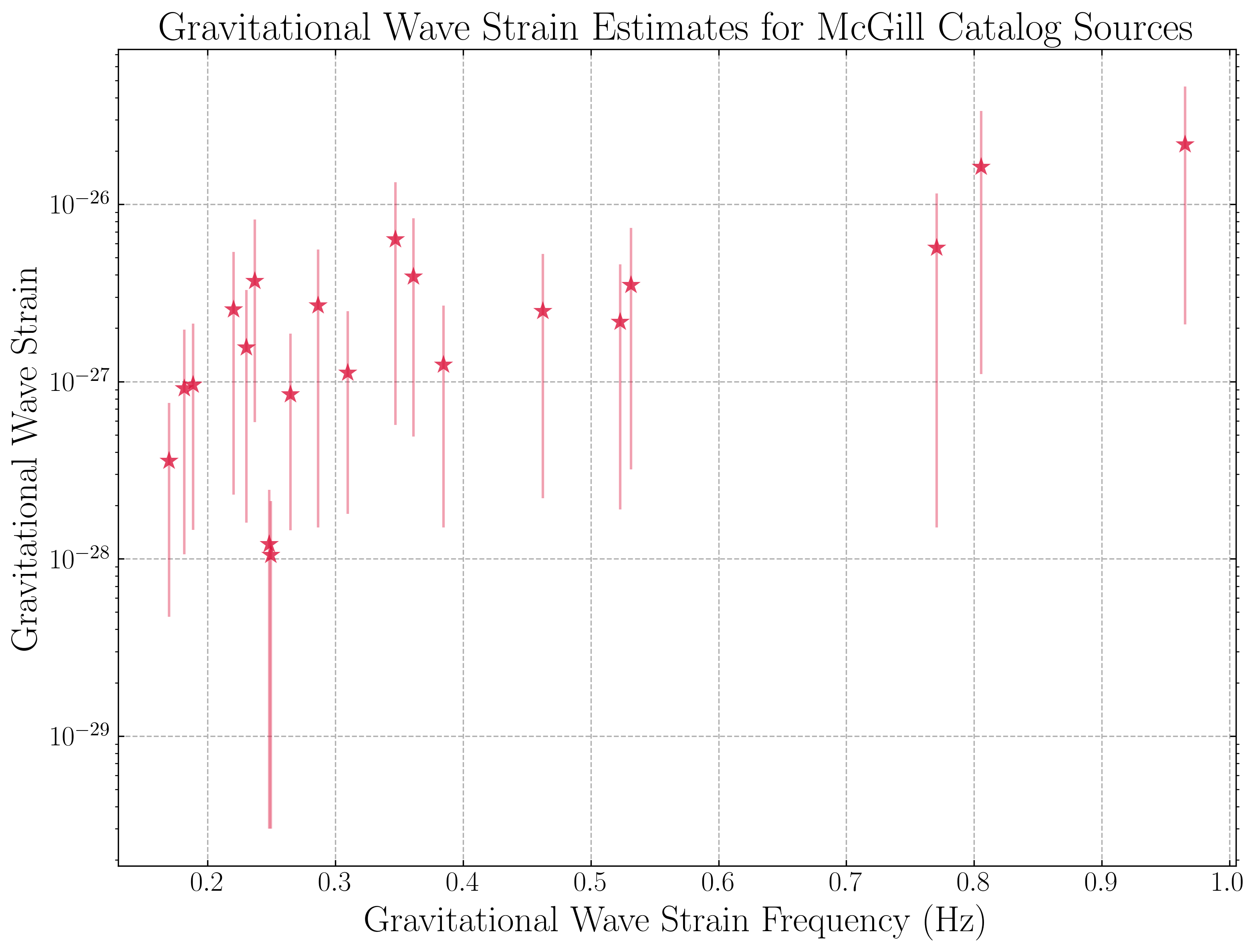}
        \caption{Wave strain estimates for sources in the McGill Magnetar Catalog computed via determined ellipticity simulations.The error margin for each source is computed for the uncertainty in the source distance. Sources in the McGill catalog not listing a specified error margin for distance are assigned a margin of $\pm 10\%$.}
        \label{fig:Mcgill_wavestrains}
\end{figure}

It is important to note that our results for magnetar ellipticity and wave strain represent upper limits for the Catalog sources.  The equation for wave strain provides upper limits to the value for the case when the rotational axis is perpendicular to the symmetry axis.  Moreover, our choice of surface field strength ($2 \times 10^{15}$ G) also results in upper limits for magnetar ellipticity and wave strain, as average surface dipolar magnetic field strength for sources in the Catalog is $\sim 3.65 \times 10^{14}$ G and the maximum detected field strength for an individual source is equal to our adopted value.

Given our goal of estimating upper limits of  wave strains for assessing the feasibility of detection of continuous gravitational waves from magnetars, simulating a reference maximum magnetic field to apply to all magnetars in the catalog is suitable for our first-order investigation. Choices in magnetic field structure and a more refined  discretized grid would be necessary to distinguish the magnetic fields, and thus ellipticities of all of the magnetars independently. Further work and far more computational resources are needed for more accurate calculations.

We compare our results qualitatively against prior gravitational wave strain predictions computed for pulsar sources. \citet{lasky_2015} compute wave strain estimates for known pulsars in the ATNF catalog, and find conventional pulsars with mixed magnetic field configurations and field strengths $|\boldsymbol{B}| < 10^{14}$ G to have strain values in the range of $\sim 10^{-34}$ to $10^{-31}$.  
We anticipate that higher magnetic field strengths will correspond to greater deformation and increased strain sensitivity magnitudes, and our results for magnetars support this reasoning and lie reasonably near past magnetar predictions.

\section{Discussion}
We implement a computational model for the stellar structure and magnetic field configuration of a magnetar to evaluate the structural changes the star undergoes as magnetic and hydrodynamic forces approach stable equilibrium. These structural changes are manifest in the principal moments of inertia which allow measurement of the stellar ellipticity. Because stellar ellipticity is derived from measurement of the principal moments of inertia via Equation \ref{eq:ellip}, damped structural evolution will limit future large-scale changes in the stellar ellipticity.

Based on these findings, we compute upper-limit estimates for ellipticity and gravitational wave strain for sources in the McGill Magnetar Catalog. In comparing our computed upper limits against prior predictions for pulsar sources \citep{Cut02, haskell_2008, Mas11, lasky_2015}, we find our results are larger than those found by other authors.  We expect that magnetars, possessing the strongest magnetic field strengths, are deformed more than conventional pulsars by their respective fields, resulting in higher ellipticities and wave strains.  Comparison of our results with other magnetar studies show that 
our ellipticity results are slightly higher than those in \cite{Gao_2017}, \cite{10.1093/mnrasl/slw072}, and \cite{10.1111/j.1365-2966.2008.12966.x}, which cite upper limits on the order of $\sim 10^{-3}$.  As discussed in \ref{sec:res_effect}, higher-resolution computations may likely bring our result to coincide with these prior works. Of note, however, is that, even with our high estimation of ellipticity, and therefore gravitational wave strain, there are no known magnetars that emit continuous gravitational waves near the sensitivity of current detectors.

In this work, we utilize the Newtonian formulation of hydrostatic equilibrium and mass conservation, which lead to analytic structural equations and a static gravitational potential. Our results provide a firm starting point for subsequent determination of magnetar wave strain upper limits, and considerable opportunity exists to extend beyond the scope of this work, including considerations for relativistic effects, dynamic gravitational potentials that evolve with the structure of the star, and adoption of a more physically representative non-barotropic EOS.

Computational limitations restricted our ability to increase the resolution of the computation for more accurate results and restricted the feasibility of increasing the total time of our simulations for rigorous stability studies. With a supercomputing allocation, this work could be extended to produce more accurate results at longer timescales. 

While these results provide valuable indication of the instrument sensitivity required to measure continuous GWs from magnetars, the operational frequency range of current GW detectors falls outside the range of frequencies produced by relatively slowly rotating magnetars. We anticipate future advancements in GW detector design to improve sensitivity to frequencies produced by magnetars, which are sure to bring about significant advancement in the scientific body of knowledge on pulsars and highly magnetic stars.



\section*{Data Availability}
The data underlying this article are available on GitHub, https://doi.org/10.5281/zenodo.4059057





\bibliographystyle{mnras}
\bibliography{main} 


\bsp	
\label{lastpage}

\newpage

\appendix

\section{Computational configuration}

Tables~\ref{tab:config} and \ref{tab:BC} lists the details of the PLUTO configurations made for the computational results presented in the paper.
\begin{table}
    \centering
    \begin{tabular}{l|c}
{\bf Setting} & {\bf Value} \\
\hline
BODY\_FORCE & POTENTIAL \\
FORCED\_TURB &  NO \\
COOLING & NO \\
RECONSTRUCTION & WENO3 \\
TIME\_STEPPING & RK3\\
DIMENSIONAL\_SPLITTING & NO\\
NTRACER & 0\\
USER\_DEF\_PARAMETERS & 0\\
EOS & IDEAL\\
ENTROPY\_SWITCH & NO\\
DIVB\_CONTROL & DIV\_CLEANING\\
BACKGROUND\_FIELD & NO\\
AMBIPOLAR\_DIFFUSION & NO\\
RESISTIVITY & NO\\
HALL\_MHD & NO\\
THERMAL\_CONDUCTION & NO\\
VISCOSITY & NO\\
ROTATING\_FRAME & NO\\
    \end{tabular}
    \caption{Configuration details for the computational stellar model.}
    \label{tab:config}
\end{table}

\begin{table}
    \centering
    \begin{tabular}{l|c}
{\bf Parameter} & {\bf Value} \\
\hline

 $r_0$ &  Outflow \\
 $r_{MAX}$ &  Outflow \\
 $\phi_0$ &  Axisymmetric \\
 $\phi_{MAX}$ & Axisymmetric \\
 $\theta_0$ & Periodic \\
 $\theta_{MAX}$ & Periodic 
    \end{tabular}
    \caption{Boundary conditions for the computation}
    \label{tab:BC}
\end{table}

\section{McGill Catalog}
Table~\ref{tab:mcgill_sources} lists all of the  magnetars in the McGill catalog.
\definecolor{Gray}{gray}{0.85}
\newcolumntype{a}{>{\columncolor{Gray}}c}

\begin{table*}
\caption{McGill catalog sources, associated attributes, and wave strain estimates. To differentiate catalog data from our wave strain findings, we place our estimates in a shaded column. We adopt the naming scheme assigned to magnetar sources by the catalog authors, including the prefix `MG' followed by the source right ascension and declination in J2000 epoch.}
\label{tab:mcgill_sources}
\begin{center}
\begin{tabular*}{\textwidth}{l@{\extracolsep{\fill}}cccca}
\hline\hline 
\rowcolor{white}
MG Name      & Distance & Period  & $f_{\text{gw}}$     & B & GW strain  \\
\rowcolor{white} & (kpc) & (s) & (Hz) & ($10^{14}$ G) & \\
\hline
MG J0100-7211 & 62.4(1.6)     & 8.020392(9)    & 0.24938 & 3.9              & 1.05$\times 10^{-28}$ \\
MG J0146+6145 & 3.6(4)      & 8.68832877(2)   & 0.2302  & 1.3              & 1.55$\times 10^{-27}$  \\
MG J0418+5372 & $\sim2$        & 9.07838822(5)   & 0.22031 & 0.061            & 2.55$\times 10^{-27}$ \\
MG J0501+4516 & $\sim2$        & 5.76209653(3)   & 0.3471  & 1.9              & 6.32$\times 10^{-27}$ \\
MG J0526-6604 & 53.6(1.2)     & 8.0544(2)   & 0.24832 & 5.6              & 1.21$\times 10^{-28}$ \\
MG J1050-5953 & 9.0(1.7)       & 6.4578754(25)  & 0.3097  & 3.9              & 1.12$\times 10^{-27}$ \\
MG J1550-5418 & 4.5(5)      & 2.0721255(1)   & 0.96525 & 3.2              & 2.17$\times 10^{-26}$ \\
MG J1622-4950 & $\sim9$        & 4.3261(1)  & 0.46231 & 2.7              & 2.49$\times 10^{-27}$  \\
MG J1635-4735 & 11.0(3)       & 2.594578(6)  & 0.77086 & 2.2              & 5.67$\times 10^{-27}$ \\
MG J1647-4552 & 3.9(7)      & 10.610644(17) & 0.18849 & <0.66             & 9.57$\times 10^{-28}$ \\
MG J1708-4008 & 3.8(5)      & 11.003027(1)  & 0.18177 & 4.6              & 9.13$\times 10^{-28}$ \\
MG J1714-3810 & $\sim$13.2     & 3.825352(4)  & 0.52283 & 5                & 2.17$\times 10^{-27}$ \\
MG J1745-2900 & $\sim$8.5      & 3.7635537(2) & 0.53141 & 1.6              & 3.49$\times 10^{-27}$ \\
MG J1808-2024 & $8.7^{+1.8}_{-1.5}$      & 7.547728(17)  & 0.26498 & 20               & 8.47$\times 10^{-28}$ \\
MG J1809-1943 & $3.5^{+0.5}_{-0.4}$     & 5.5403537(2)  & 0.36099 & 2.1              & 3.91$\times 10^{-27}$  \\
MG J1822-1604 & 1.6(3)      & 8.43771958(6)  & 0.23703 & 0.51             & 3.69$\times 10^{-27}$ \\
MG J1833-0831 & ... & 7.5654084(4) & 
0.26436 & 1.6 & ...\\
MG J1834-0845 & 4.2(3)      & 2.4823018(1)  & 0.8057  & 1.4              & 1.62$\times 10^{-26}$ \\
MG J1841-0456 & $8.5^{+1.3}_{-1.0}$      & 11.782898(1) & 0.16974 & 6.9              & 3.56$\times 10^{-28}$ \\
MG J1907+0919 & 12.5(1.7)     & 5.19987(7)  & 0.38463 & 7                & 1.24$\times 10^{-27}$ \\
MG J2301+5852 & 3.2(2)      & 6.978948446(4)  & 0.28658 & 0.59             & 2.69$\times 10^{-27}$  \\ \hline
\end{tabular*}
\end{center}
\end{table*}



\end{document}